\RequirePackage{lineno}
\documentclass[aps,prl,twocolumn,showpacs,superscriptaddress,groupedaddress]{revtex4}

 \usepackage{CJK}
\usepackage{graphicx} % Include figure files
\usepackage{color}
\usepackage{dcolumn}  % Align table columns on decimal point
\usepackage{tabularx,array}

\newcolumntype{C}{>{\centering\arraybackslash}X}
\usepackage{multirow}
\RequirePackage{hyperref}
\hypersetup{colorlinks=false}

\usepackage{amsmath}

\graphicspath{{ps}}

\begin{document}

\widetext

%\resizebox{!}{3cm}{\includegraphics{belle.eps}}

% \rightline{\vbox{ \hbox{Belle DRAFT FINAL (3)}
%                   \hbox{Intended for {\it PRL}}
%                   \hbox{Author: C.-W. Lin, Y.-C. Chen,} \hbox{Y.-J. Lee, P. Chang}
%                   \hbox{Committee: R.Seidl(chair),}
%                   \hbox{W. Jacobs, M. Niiyama}
% }}

% \linenumbers
%%% Talk title
\title{ \quad\\[0.5cm]  Measurement of Two-Particle Correlations of Hadrons in $e^{+}e^{-}$ Collisions at Belle}

%%% Paper:    Two-particle correlations in e+e- collisions
%%% Journal:  Physical Review Letters
%%% Contacts: C.-W. Lin (willylin@hep1.phys.ntu.edu.tw)
%%%           Y.-C. Chen (janice.chen45@gmail.com)
%%%           Y.-J. Lee (yen-jie.lee@cern.ch)
%%%           P. Chang (pchang@phys.ntu.edu.tw)
%%% Non-responding authors or those who said NO are commented out.
%%% ====================================================================
%%% Click the RELOAD button on your web browser to see the updated file.
%%% ====================================================================
%%% Use \input{author} to insert this material into your latex file.
%%%%% Force institutions to appear in alphabetical order when typeset.
\noaffiliation
\affiliation{Department of Physics, University of the Basque Country UPV/EHU, 48080 Bilbao}
\affiliation{University of Bonn, 53115 Bonn}
\affiliation{Brookhaven National Laboratory, Upton, New York 11973}
\affiliation{Budker Institute of Nuclear Physics SB RAS, Novosibirsk 630090}
\affiliation{Faculty of Mathematics and Physics, Charles University, 121 16 Prague}
%%%\affiliation{Chiba University, Chiba 263-8522}
%%%\affiliation{Chonnam National University, Gwangju 61186}
\affiliation{Chung-Ang University, Seoul 06974}
\affiliation{University of Cincinnati, Cincinnati, Ohio 45221}
\affiliation{Deutsches Elektronen--Synchrotron, 22607 Hamburg}
\affiliation{Duke University, Durham, North Carolina 27708}
\affiliation{Institute of Theoretical and Applied Research (ITAR), Duy Tan University, Hanoi 100000}
\affiliation{University of Florida, Gainesville, Florida 32611}
\affiliation{Department of Physics, Fu Jen Catholic University, Taipei 24205}
\affiliation{Key Laboratory of Nuclear Physics and Ion-beam Application (MOE) and Institute of Modern Physics, Fudan University, Shanghai 200443}
\affiliation{Justus-Liebig-Universit\"at Gie\ss{}en, 35392 Gie\ss{}en}
\affiliation{Gifu University, Gifu 501-1193}
%%%\affiliation{II. Physikalisches Institut, Georg-August-Universit\"at G\"ottingen, 37073 G\"ottingen}
\affiliation{SOKENDAI (The Graduate University for Advanced Studies), Hayama 240-0193}
%%%\affiliation{Gyeongsang National University, Jinju 52828}
\affiliation{Department of Physics and Institute of Natural Sciences, Hanyang University, Seoul 04763}
\affiliation{University of Hawaii, Honolulu, Hawaii 96822}
\affiliation{High Energy Accelerator Research Organization (KEK), Tsukuba 305-0801}
\affiliation{J-PARC Branch, KEK Theory Center, High Energy Accelerator Research Organization (KEK), Tsukuba 305-0801}
\affiliation{National Research University Higher School of Economics, Moscow 101000}
\affiliation{Forschungszentrum J\"{u}lich, 52425 J\"{u}lich}
%%%\affiliation{Hiroshima University, Higashi-Hiroshima, Hiroshima 739-8530}
\affiliation{IKERBASQUE, Basque Foundation for Science, 48013 Bilbao}
\affiliation{Indian Institute of Science Education and Research Mohali, SAS Nagar, 140306}
%%%\affiliation{Indian Institute of Technology Bhubaneswar, Satya Nagar 751007}
\affiliation{Indian Institute of Technology Guwahati, Assam 781039}
\affiliation{Indian Institute of Technology Hyderabad, Telangana 502285}
\affiliation{Indian Institute of Technology Madras, Chennai 600036}
\affiliation{Indiana University, Bloomington, Indiana 47408}
%%%\affiliation{Institute of High Energy Physics, Chinese Academy of Sciences, Beijing 100049}
\affiliation{Institute of High Energy Physics, Vienna 1050}
\affiliation{Institute for High Energy Physics, Protvino 142281}
%%%\affiliation{Institute of Mathematical Sciences, Chennai 600113}
\affiliation{INFN - Sezione di Napoli, I-80126 Napoli}
\affiliation{INFN - Sezione di Roma Tre, I-00146 Roma}
\affiliation{INFN - Sezione di Torino, I-10125 Torino}
\affiliation{Iowa State University, Ames, Iowa 50011}
\affiliation{Advanced Science Research Center, Japan Atomic Energy Agency, Naka 319-1195}
\affiliation{J. Stefan Institute, 1000 Ljubljana}
%%%\affiliation{Kanagawa University, Yokohama 221-8686}
\affiliation{Institut f\"ur Experimentelle Teilchenphysik, Karlsruher Institut f\"ur Technologie, 76131 Karlsruhe}
\affiliation{Kavli Institute for the Physics and Mathematics of the Universe (WPI), University of Tokyo, Kashiwa 277-8583}
%%%\affiliation{King Abdulaziz City for Science and Technology, Riyadh 11442}
\affiliation{Department of Physics, Faculty of Science, King Abdulaziz University, Jeddah 21589}
\affiliation{Kitasato University, Sagamihara 252-0373}
\affiliation{Korea Institute of Science and Technology Information, Daejeon 34141}
\affiliation{Korea University, Seoul 02841}
%%%\affiliation{Kyoto Sangyo University, Kyoto 603-8555}
\affiliation{Kyungpook National University, Daegu 41566}
%%%\affiliation{Universit\'{e} Paris-Saclay, CNRS/IN2P3, IJCLab, 91405 Orsay}
%%%\affiliation{\'Ecole Polytechnique F\'ed\'erale de Lausanne (EPFL), Lausanne 1015}
\affiliation{P.N. Lebedev Physical Institute of the Russian Academy of Sciences, Moscow 119991}
%%%\affiliation{Liaoning Normal University, Dalian 116029}
\affiliation{Faculty of Mathematics and Physics, University of Ljubljana, 1000 Ljubljana}
\affiliation{Ludwig Maximilians University, 80539 Munich}
\affiliation{Luther College, Decorah, Iowa 52101}
\affiliation{Malaviya National Institute of Technology Jaipur, Jaipur 302017}
%%%\affiliation{University of Malaya, 50603 Kuala Lumpur}
\affiliation{Faculty of Chemistry and Chemical Engineering, University of Maribor, 2000 Maribor}
\affiliation{Max-Planck-Institut f\"ur Physik, 80805 M\"unchen}
\affiliation{School of Physics, University of Melbourne, Victoria 3010}
\affiliation{University of Mississippi, University, Mississippi 38677}
\affiliation{University of Miyazaki, Miyazaki 889-2192}
%%%\affiliation{Moscow Physical Engineering Institute, Moscow 115409}
\affiliation{Graduate School of Science, Nagoya University, Nagoya 464-8602}
\affiliation{Kobayashi-Maskawa Institute, Nagoya University, Nagoya 464-8602}
\affiliation{Universit\`{a} di Napoli Federico II, I-80126 Napoli}
\affiliation{Nara Women's University, Nara 630-8506}
%%%\affiliation{National Central University, Chung-li 32054}
\affiliation{National United University, Miao Li 36003}
\affiliation{Department of Physics, National Taiwan University, Taipei 10617}
%%%\affiliation{H. Niewodniczanski Institute of Nuclear Physics, Krakow 31-342}
\affiliation{Nippon Dental University, Niigata 951-8580}
\affiliation{Niigata University, Niigata 950-2181}
%%%\affiliation{University of Nova Gorica, 5000 Nova Gorica}
\affiliation{Novosibirsk State University, Novosibirsk 630090}
%%%\affiliation{Okinawa Institute of Science and Technology, Okinawa 904-0495}
\affiliation{Osaka City University, Osaka 558-8585}
\affiliation{Pacific Northwest National Laboratory, Richland, Washington 99352}
\affiliation{Panjab University, Chandigarh 160014}
%%%\affiliation{Peking University, Beijing 100871}
\affiliation{University of Pittsburgh, Pittsburgh, Pennsylvania 15260}
\affiliation{Punjab Agricultural University, Ludhiana 141004}
%%%\affiliation{Research Center for Electron Photon Science, Tohoku University, Sendai 980-8578}
\affiliation{Research Center for Nuclear Physics, Osaka University, Osaka 567-0047}
\affiliation{Meson Science Laboratory, Cluster for Pioneering Research, RIKEN, Saitama 351-0198}
%%%\affiliation{Theoretical Research Division, Nishina Center, RIKEN, Saitama 351-0198}
\affiliation{RIKEN BNL Research Center, Upton, New York 11973}
\affiliation{Dipartimento di Matematica e Fisica, Universit\`{a} di Roma Tre, I-00146 Roma}
\affiliation{Department of Modern Physics and State Key Laboratory of Particle Detection and Electronics, University of Science and Technology of China, Hefei 230026}
\affiliation{Seoul National University, Seoul 08826}
\affiliation{Showa Pharmaceutical University, Tokyo 194-8543}
%%%\affiliation{Soochow University, Suzhou 215006}
\affiliation{Soongsil University, Seoul 06978}
%%%\affiliation{University of South Carolina, Columbia, South Carolina 29208}
%%%\affiliation{Stefan Meyer Institute for Subatomic Physics, Vienna 1090}
\affiliation{Sungkyunkwan University, Suwon 16419}
\affiliation{School of Physics, University of Sydney, New South Wales 2006}
\affiliation{Department of Physics, Faculty of Science, University of Tabuk, Tabuk 71451}
\affiliation{Tata Institute of Fundamental Research, Mumbai 400005}
%%%\affiliation{Excellence Cluster Universe, Technische Universit\"at M\"unchen, 85748 Garching}
\affiliation{Department of Physics, Technische Universit\"at M\"unchen, 85748 Garching}
%%%\affiliation{School of Physics and Astronomy, Tel Aviv University, Tel Aviv 69978}
\affiliation{Toho University, Funabashi 274-8510}
\affiliation{Department of Physics, Tohoku University, Sendai 980-8578}
\affiliation{Earthquake Research Institute, University of Tokyo, Tokyo 113-0032}
\affiliation{Department of Physics, University of Tokyo, Tokyo 113-0033}
\affiliation{Tokyo Institute of Technology, Tokyo 152-8550}
%%%\affiliation{Tokyo Metropolitan University, Tokyo 192-0397}
%%%\affiliation{Utkal University, Bhubaneswar 751004}
\affiliation{Virginia Polytechnic Institute and State University, Blacksburg, Virginia 24061}
\affiliation{Wayne State University, Detroit, Michigan 48202}
%%%\affiliation{Yamagata University, Yamagata 990-8560}
\affiliation{Yonsei University, Seoul 03722}
  \author{Y.-C.~Chen}\thanks{now at Massachusetts Institute of Technology}\affiliation{Department of Physics, National Taiwan University, Taipei 10617} % Taiwan
\author{Y.-J.~Lee}\thanks{now at Massachusetts Institute of Technology}\affiliation{Department of Physics, National Taiwan University, Taipei 10617} % Taiwan
  \author{P.~Chang}\affiliation{Department of Physics, National Taiwan University, Taipei 10617} % Taiwan
  \author{I.~Adachi}\affiliation{High Energy Accelerator Research Organization (KEK), Tsukuba 305-0801}\affiliation{SOKENDAI (The Graduate University for Advanced Studies), Hayama 240-0193} % KEK
% \author{K.~Adamczyk}\affiliation{H. Niewodniczanski Institute of Nuclear Physics, Krakow 31-342} % Krakow
% \author{J.~K.~Ahn}\affiliation{Korea University, Seoul 02841} % Korea
  \author{H.~Aihara}\affiliation{Department of Physics, University of Tokyo, Tokyo 113-0033} % Tokyo
  \author{S.~Al~Said}\affiliation{Department of Physics, Faculty of Science, University of Tabuk, Tabuk 71451}\affiliation{Department of Physics, Faculty of Science, King Abdulaziz University, Jeddah 21589} % Tabuk
  \author{D.~M.~Asner}\affiliation{Brookhaven National Laboratory, Upton, New York 11973} % BNL
% \author{H.~Atmacan}\affiliation{University of Cincinnati, Cincinnati, Ohio 45221} % Cincinnati
% \author{V.~Aulchenko}\affiliation{Budker Institute of Nuclear Physics SB RAS, Novosibirsk 630090}\affiliation{Novosibirsk State University, Novosibirsk 630090} % BINP
  \author{T.~Aushev}\affiliation{National Research University Higher School of Economics, Moscow 101000} % HSE
  \author{R.~Ayad}\affiliation{Department of Physics, Faculty of Science, University of Tabuk, Tabuk 71451} % Tabuk
% \author{T.~Aziz}\affiliation{Tata Institute of Fundamental Research, Mumbai 400005} % Tata
  \author{V.~Babu}\affiliation{Deutsches Elektronen--Synchrotron, 22607 Hamburg} % DESY
% \author{S.~Bahinipati}\affiliation{Indian Institute of Technology Bhubaneswar, Satya Nagar 751007} % IITB
% \author{A.~M.~Bakich}\affiliation{School of Physics, University of Sydney, New South Wales 2006} % Sydney
% \author{Y.~Ban}\affiliation{Peking University, Beijing 100871} % Peking
% \author{E.~Barberio}\affiliation{School of Physics, University of Melbourne, Victoria 3010} % Melbourne
% \author{M.~Barrett}\affiliation{High Energy Accelerator Research Organization (KEK), Tsukuba 305-0801} % KEK
% \author{M.~Bauer}\affiliation{Institut f\"ur Experimentelle Teilchenphysik, Karlsruher Institut f\"ur Technologie, 76131 Karlsruhe} % Karlsruhe
% ...
% Lebedev
% \author{V.~Zhulanov}\affiliation{Budker Institute of Nuclear Physics SB RAS, Novosibirsk 630090}\affiliation{Novosibirsk State University, Novosibirsk 630090} % BINP
\collaboration{The Belle Collaboration}

% \collaboration{The Belle Collaboration}
\date{\today}

\noaffiliation

\begin{abstract}
% \linenumbers
The measurement of two-particle angular correlation functions in high-multiplicity $e^+e^-$ collisions at $\sqrt{s}=10.52$ GeV is reported. 
In this study, the $89.5~{\rm fb}^{-1}$ of hadronic $e^+e^-$ annihilation data collected by the Belle detector at KEKB are used.
Two-particle angular correlation functions are measured in the full relative azimuthal angle ($\Delta \phi$) and three units of pseudorapidity ($\Delta \eta$), defined by either the electron beam axis or the event-shape thrust axis, and are studied as a function of charged-particle multiplicity.
The measurement in the thrust axis analysis,
with mostly outgoing quark pairs determining the reference axis, is sensitive to the region of additional soft gluon emissions.
No significant anisotropic collective behavior is observed with either coordinate analyses.
Near-side jet correlations appear to be absent in the thrust axis analysis.
The measurements are compared to predictions from various event generators and are expected to provide new constraints to the phenomenological models in the low-energy regime.

\end{abstract}

% https://ufn.ru/en/pacs/

\maketitle

%%%% >>>> keep the final version single-spaced
\tighten

{\renewcommand{\thefootnote}{\fnsymbol{footnote}}}
\setcounter{footnote}{0}

Two-particle angular correlations have been extensively studied in search of quark-gluon plasma (QGP) formation and its properties in nucleus-nucleus collisions~\cite{Arsene:2004fa,Back:2004je,Adams:2005dq,Adcox:2004mh} over the last several decades. 
In such collisions, a ridge-like structure of the correlation function, residing in a particular phase space where particle pairs have large differences in pseudorapidity but small differences in azimuthal angle, is observed.
This signal in relativistic heavy ion collisions is interpreted as the macroscopic consequence of the hydrodynamical expansion of the perfect-fluid-like QGP state with the presence of initial density fluctuations~\cite{Ollitrault:1992bk,Alver:2010gr,Heinz:2013th}.
The ridge-like signal was also observed in high charged-particle multiplicity events in proton-proton, proton-nucleus, deuteron-nucleus, and helium-nucleus collisions~\cite{Khachatryan:2010gv,Aad:2015gqa,CMS:2012qk,Abelev:2012ola,Aad:2012gla,Adare:2013piz,Chatrchyan:2012wg,Aaij:2015qcq,Adam:2019woz}. 
Recently, data on ultraperipheral PbPb photonuclear collisions~\cite{ATLAS:2021jhn} also resulted in significant second- and third-order flow coefficients, which is an approach to quantify the two-particle azimuthal anisotropy with Fourier harmonics. Essentially, the ridge-like signal is reported in all collision systems involving at least one hadron.
However, the physical origin of azimuthal anisotropies in these smaller collision systems is not yet fully understood~\cite{Dusling:2015gta,Nagle:2018nvi}. In hadron-hadron collisions, the complexity introduced by the initial state cannot be easily factored out. A large number of theoretical models based on different underlying mechanisms such as partonic initial-state correlations~\cite{Dusling:2013qoz}, final-state interactions~\cite{He:2015hfa,Bierlich:2017vhg}, and hydrodynamic medium expansion~\cite{Bozek:2011if} have been proposed to explain the observed ridge-like signal in these small systems. 

To break down the question, high charged-particle multiplicity events produced in the even smaller electron-ion and  electron-positron collision systems are proposed to provide accessibility to understanding the cause of this special collective behavior~\cite{Nagle:2017sjv}.
As an example, a color dipole configuration (two color strings aligned in parallel with a gap in between) in the $e^+e^-$ collision system can exhibit anisotropy in the initial parton geometry and generate ridge-like correlations. 
Recently, experimental studies have been extended to such smaller collision systems, e.g., electron-proton~\cite{ZEUS:2019jya} and electron-positron ($e^+e^-$)~\cite{Badea:2019vey} collisions. No significant ridge-like signal was observed in these measurements. 
These results have stimulated discussions on the ways to search for and understand possible collectivity signatures
in $e^+e^-$~\cite{Bierlich:2020naj,Castorina:2020iia,Baty:2021ugw} and electron-ion collisions~\cite{Shi:2020djm,Agostini:2020fmq}. 
However, the data samples used for the search in the $e^+e^-$ ALEPH archived data~\cite{Badea:2019vey} is small, which motivates the examination of a high-statistics data to study the highest multiplicity tail at Belle.

Taking advantage of the clean environment in $e^+e^-$ collisions and high-statistics data collected with the Belle detector at KEKB~\cite{KEKBCitation}, the analysis is performed for the first time at a center-of-mass energy of $\sqrt{s}=10.52$ GeV, which is $60$ MeV lower than the $\Upsilon(4S)$ resonance. Overall, a data sample of $89.5~{\rm fb}^{-1}$ is utilized in this analysis, which is the full dataset of collisions at $\sqrt{s}=10.52$~GeV.
This analysis closely follows the previous analysis procedure with ALEPH archived data~\cite{Badea:2019vey}.
Although the average event multiplicity is lower than the ALEPH data, the two-particle correlation analysis is performed on the largest off-resonance Belle dataset, whose results can solidify previous findings.
The Belle hadronic-event dataset is about four times larger than that with ALEPH archived data, enabling the collectivity search to move forward from a scan amongst the $0.5\%$ highest multiplicity events of the total distribution to that of $0.02\%$ percentile.
Moreover, the measurement of the two-particle correlation function in the low-energy regime can provide additional inputs to the phenomenological fragmentation models.

The Belle detector is a large-solid-angle magnetic spectrometer that consists of a silicon vertex detector, a 50-layer central drift chamber, an array of aerogel threshold Cherenkov counters, a barrel-like arrangement of time-of-flight scintillation counters, and an electromagnetic calorimeter (ECL) comprising CsI(Tl) crystals located inside a superconducting solenoid coil that provides a 1.5~T magnetic field.  An iron flux-return located outside of the coil is instrumented to detect $K_L^0$ mesons and muons.  The detector is described in detail elsewhere~\cite{BelleDetectorCitation}.

% event selection
The hadronic-event selection~\cite{Belle:2001jqo}, including requirements on event multiplicity and energy sum in the ECL, is adopted to suppress contamination from two-photon, radiative Bhabha, and other QED events.
% track selection
Particles used in the calculation of the correlation functions are primary charged  tracks, defined as prompt tracks or decay products of intermediate particles with proper lifetime $\tau<1 {\rm~cm}/c$. The corresponding selection on experimental data is tracks produced from the interaction point (including from short-lifetime particle decays) and tracks from long-lifetime particle decays with $V_{r}< 1\rm~cm$, where $V_{r}$ is the distance in the transverse plane of the decay vertex from the interaction point. The latter can intersect with other track on a common space-point and has dihadron invariant mass consistent with the mass of a $K_S^0$ ($0.480$-$0.516$~GeV$/c^2$) or $\Lambda^0/\bar{\Lambda}^0$ ($1.111$-$1.121$~GeV$/c^2$) candidate; otherwise, the track is deemed as a prompt track and accepted in the primary particle selection.
Charged tracks are required to be within the detector acceptance, i.e., with polar angles ranging from $17^\circ$-$150^\circ$ ($+z$ is defined opposite to the $e^+$ beam), and have transverse momenta in the center-of-mass frame greater than $0.2$~GeV$/c$. 
Impact parameter requirements are adopted to select charged tracks within $\pm 2$ cm with respect to the interaction point in the transverse plane, and $\pm 5$ cm along the $z$ direction.
For a pair of neighboring low-momentum tracks with the absolute value of cosine opening angle greater than $0.95$ and transverse momenta below $0.4$~GeV$/c$, the one with less momentum is deemed as a duplicated track and is hence removed. Tracks from photon conversion candidates are vetoed, with the latter selected with the invariant mass less than $0.25$~GeV$/c^2$ and the decay-vertex radius greater than $1.5$~cm.
% Detailed selections will be documented in Ref.~\cite{ToBePublished}.

%%%%%%%%%%%%%%%%%%%% Efficiency correction %%%%%%%%%%%%%%%%%%%%
To eliminate the effects of the nonuniform detection efficiency and misreconstruction bias, efficiency correction factors are derived by the Belle Monte Carlo (MC) sample. 
The Belle MC is simulated based on \textsc{evtgen}~\cite{Ryd:2005zz} and \textsc{pythia6}~\cite{2001CoPhC.135..238S}, where hadronic $q\bar{q}$ ($q=u,d,s,c$) fragmentation as well as low-multiplicity $e^{+}e^{-} \to \tau^{+}\tau^{-}$ and two-photon processes are taken into account.
The detector response is simulated with GEANT3~\cite{Brun:1119728}.
%%%%%%%%%%%%%%%%%%%% MC Reweighting %%%%%%%%%%%%%%%%%%%%
The MC sample is further reweighted to match event multiplicity and thrust distributions of the data in order to correct for the imperfection in MC simulation.
%%%%%%%%%%%%%%%%%%%% NTRK CLASS %%%%%%%%%%%%%%%%%%%%
In order to study the multiplicity dependence of the correlation function, the events are binned into multiplicity classes using the reconstructed track multiplicity, denoted ${\rm N}_{\rm trk}^{\rm rec}$, by counting tracks after all selections.
For low-multiplicity events with ${\rm N}_{\rm trk}^{\rm rec}<12$, only a sample size of $11.5~{\rm fb}^{-1}$ is used.
The multiplicity classes used in this study, their corresponding fraction of data, and the mapping of average reconstructed multiplicities $\langle {\rm N}_{\rm trk}^{\rm rec}\rangle$ to average multiplicities after efficiency correction $\langle {\rm N}_{\rm trk}^{\rm corr}\rangle$ are listed in Table~\ref{tab:NtrkCorr}.

\begin{table}[ht]
\begin{center}
\caption{Average multiplicities and corrected multiplicities of different ${\rm N}_{\rm trk}^{\rm rec}$ intervals. 
}
\begin{tabularx}{0.46\textwidth}{
	>{\centering}p{0.13\textwidth}
	>{\centering}p{0.17\textwidth}
	>{\centering}p{0.08\textwidth}
	C
}
\hline \hline
${\rm N}_{\rm trk}^{\rm rec}$ range & Fraction of data (\%) & $\langle {\rm N}_{\rm trk}^{\rm rec}\rangle$ & $\langle {\rm N}_{\rm trk}^{\rm corr}\rangle$\\
\hline 
$[6, 10)$       & 44.33 &  6.98 & 7.05 \\
$[10, 12)$      &  2.65 & 10.26 & 10.12\\
$[12, 14)$      &  0.29 & 12.20 & 11.90\\
% $[12, \infty)$  &  0.30 & 12.31 & 12.03\\
$[14, \infty)$  &  0.02 & 14.22 & 14.24\\
\hline \hline
\end{tabularx} 
\label{tab:NtrkCorr}
\end{center}
\end{table}

%%%%%%%%%%%%%%%%%%%% 2PC METHODOLOGY  %%%%%%%%%%%%%%%%%%%%
The Belle experiment is operated with the KEKB asymmetric energy collider, colliding the $8$~GeV electron beam and the $3.5$~GeV positron beam. We boost events to their center-of-mass frame to perform the angular correlation analysis.
Following a procedure similar to what has already been established in Ref.~\cite{CMS:2012qk}, the two-particle correlation function observable is defined as

\begin{equation}
\frac{1}{{\rm N}_{\rm trk}^{\rm corr}}\frac{d^2{\rm N}^{\rm pair}}{d\Delta\eta d\Delta\phi}= B(0,0) \times \frac{S(\Delta\eta, \Delta\phi)}{B(\Delta\eta, \Delta\phi)}.
\label{eqn:2PCAssociatedYield}
\end{equation}
The correlation function is expressed in terms of the particle pair's angular difference $\Delta\eta = \pm (\eta_i - \eta_j)$ and $\Delta\phi = \pm (\phi_i - \phi_j)$, where $i,j$ label the track pair's indices. The calculation is based on an assumption of correlations being symmetric about the origin $(\Delta \eta, \Delta \phi) = (0,0)$; hence, four entries are counted for a given pair.
The per-charged-particle associated track-pair yield is denoted as ${\rm N}^{\rm pair}$, and is reweighted by efficiency correction factors of both particles.
The signal correlation $S(\Delta\eta, \Delta\phi)$ and the background correlation $B(\Delta\eta, \Delta\phi)$ can be explicitly written out as

\begin{equation}
\begin{aligned}
S(\Delta\eta, \Delta\phi) &= \frac{1}{{\rm N}_{\rm trk}^{\rm corr}}\frac{d^2{\rm N}^{\rm same}} {d\Delta\eta d\Delta\phi},\\
B(\Delta\eta, \Delta\phi) &= \frac{1}{{\rm N}_{\rm trk}^{\rm corr}}\frac{d^2{\rm N}^{\rm mix}}{d\Delta\eta d\Delta\phi},
\end{aligned}
\label{eqn:2PCSigBkgFunctions}
\end{equation}
where ${\rm N}^{\rm same}$ (${\rm N}^{\rm mix}$) counts the number of track-pairs formed by matching the $i$-th charged particle of a given event with the $j$-th particle in the same event (``mixed event''~\cite{Khachatryan:2010gv}).
A mixed event in this work is a combination of tracks from three random events and is normalized by a factor of $1/3$. Three random events are chosen such that their ${\rm N}_{\rm trk}^{\rm rec}$'s are the same as that of the event they match to.
The $B(0,0)$ factor is incorporated in the calculation of the correlation function, serving as the normalization of the artificially constructed $B(\Delta\eta, \Delta\phi)$. This factor is obtained by extrapolating the function value to the origin of $B(\Delta\eta, \Delta\phi)$.
An additional correction on the correlation function is applied to deal with the effects introduced by using finite-bin histogramming to approximate the density function.
The bin-size effect is modeled by a second-order polynomial and the magnitude of the correlation function is calibrated. 
%%%%%%%%%%%%%%%%%%%% Residual reconstructed effect correction: Bin-by-bin %%%%%%%%%%%%%%%%%%%%
To unfold back to the truth level, final correlations are corrected with the bin-by-bin method~\cite{Choudalakis:2011rr}, accounting for residual reconstruction effects after efficiency corrections. 
% Detailed correction procedures will be provided in Ref.~\cite{ToBePublished}.
 
%%%%%%%%%%%%%%%%%%%% INTRO OF THRUST AXIS ANALYSIS  %%%%%%%%%%%%%%%%%%%%
The two-particle correlation function is explored in two coordinate systems: beam and thrust axis coordinates in the $e^+e^-$ center-of-mass frame. The former is the same
as that presented in most of the two-particle correlation studies, while in the latter, initiated by Ref.~\cite{Badea:2019vey},  the event thrust axis~\cite{PhysRevLett.39.1587} is used as the reference axis,
with missing momentum of the event included. 
%%% reweight mix
The construction of mixed events in the thrust axis analysis is identical to that in the beam axis analysis, requiring the multiplicity matching only.
In the thrust axis coordinate analysis, the kinematics ($p_{\rm T}, \eta, \phi$) of a mixed event are calculated with respect to the thrust axis of its matched physical event.  To adjust the kinematics distribution of the mixed event to physical kinematic ($p_{\rm T}, \eta, \phi$) spectra, a reweighting correction is adopted. 
% The full thrust axis analysis construction will be documented in Ref.~\cite{ToBePublished}.

In the $e^{+} e^{-}$ annihilation process, when the interacting system is located in between or along the color string connecting the $q\bar{q}$ pair, measuring with a coordinate system defined by the event thrust axis provides a more direct picture. 
From the viewpoint of relativistic fluid dynamics~\cite{Heinz:2013th}, conventional beam-axis measurements are sensitive 
to features within the plane transverse to the collision axis,
% in their transverse directions
probing any anisotropic behavior of the QCD medium, which are widely studied as the phenomena of elliptic or triangular flow~\cite{Thomas:2010zym, Alver:2010gr,PhysRevLett.105.252302}.
The insensitive region of the two-particle correlation function in the beam axis analysis is at the beam pipe, where a particle pair with a large pseudorapidity difference is excluded from the finite $\Delta \eta$ region of interest (e.g., $|\Delta \eta| \le 3.0$ in this analysis).
In addition, the on-axis track-pair correlation is too deformed to form an obvious correlation structure, since the $\phi$ coordinate is ill-represented near both poles of the spherical coordinate. 
Correspondingly, the insensitivity of the thrust-axis correlation function is at its reference thrust axis which quark-initiated dijets are close to; however, they are sensitive in the mid-rapidity region, where additional soft gluons emit apart from the leading quark-antiquark dijet-like structure. The sensitivity to the finer structure allows one to check in details if there are special correlations among the color activity in the small system.

\begin{figure}[ht]
\centering
		\includegraphics[width=0.40\textwidth]{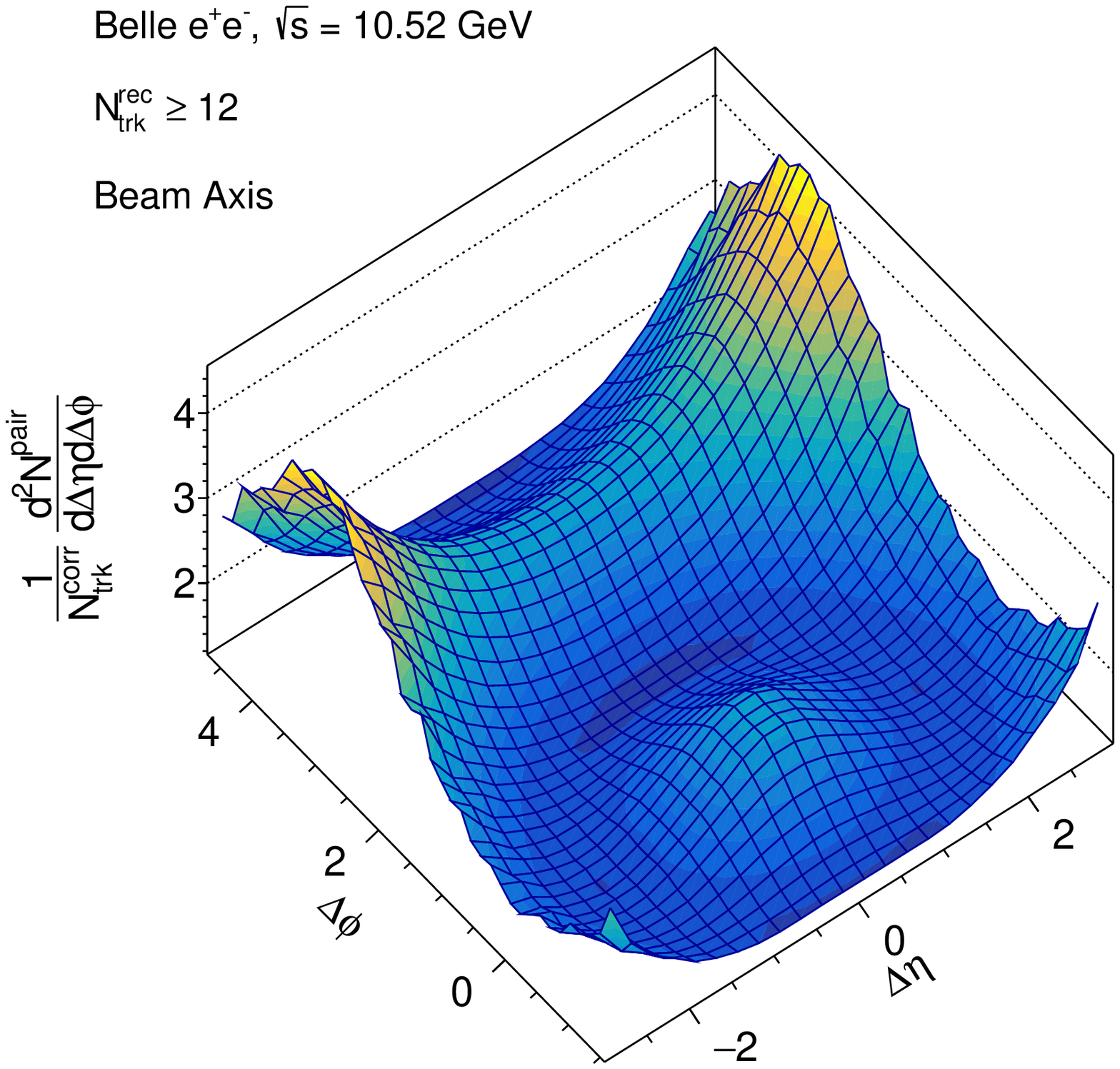}
		\includegraphics[width=0.40\textwidth]{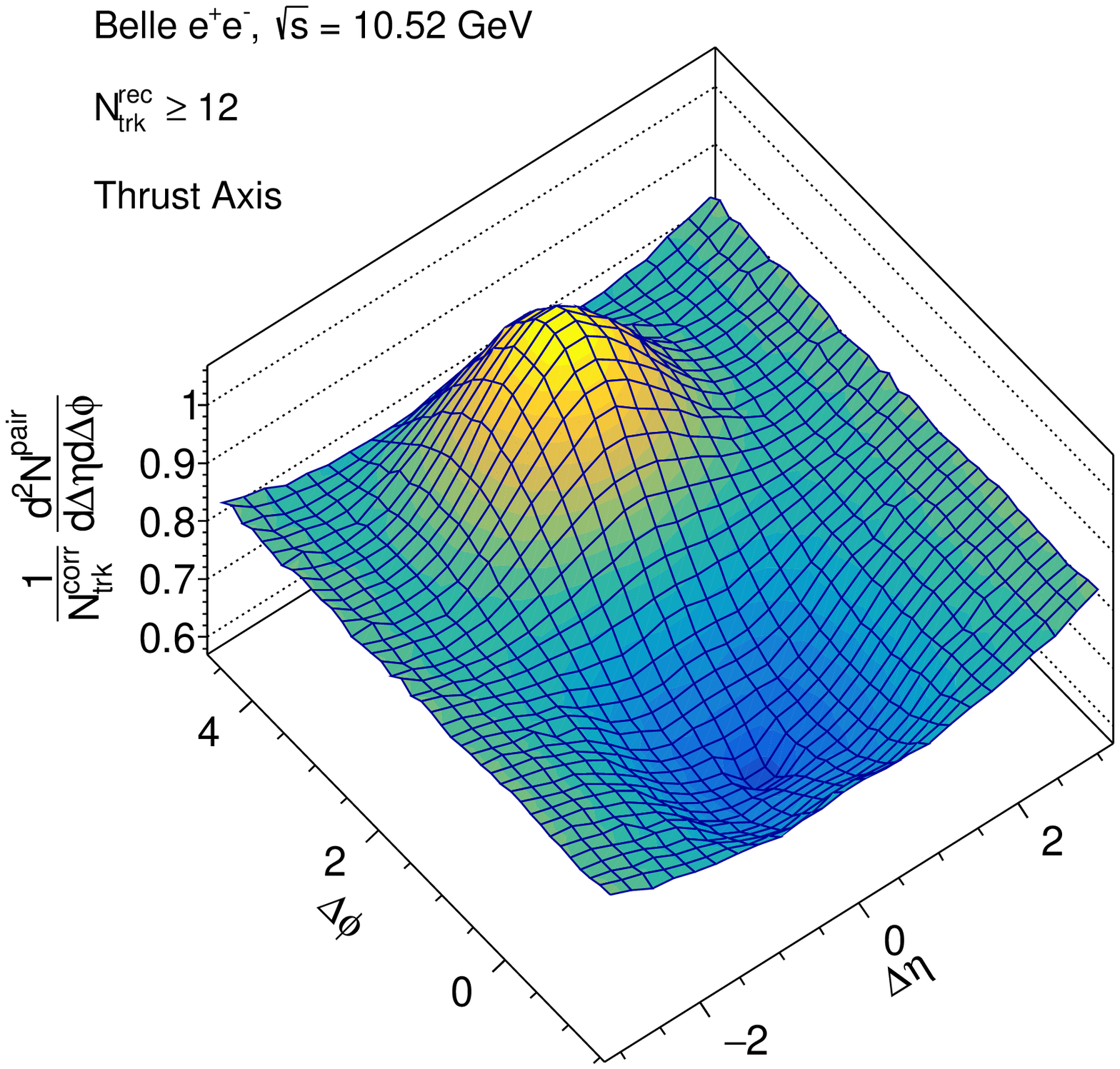}
\caption{Two-particle correlation functions for beam (top) and thrust (bottom) axis analyses with the multiplicity ${\rm N}_{\rm trk}^{\rm rec} \ge 12$.}
\label{fig:2PC}
\end{figure}

%%%%%%%%%%%%%%%%%%%% 2PC RESULT %%%%%%%%%%%%%%%%%%%%
In Fig.~\ref{fig:2PC}, correlation functions 
with multiplicity ${\rm N}_{\rm trk}^{\rm rec} \ge 12$ are shown for both beam and thrust axis coordinates.
In the beam axis coordinate view,
the peak near the origin $(\Delta\eta, \Delta\phi)=(0,0)$ has contributions from pairs originating in the same jet, while the structure at $\Delta\phi\approx\pi$ is from back-to-back correlations. These features reflect the two-particle correlation of dijet-like $q \bar{q}$ events, which mainly contribute in $e^+e^-$ collisions.
% thrust axis analysis
In contrast, for the thrust axis coordinates, the dominant structure is the hill-like bump near $(\Delta\eta, \Delta\phi)\approx(0,\pi)$, 
while a sizeable near-side correlation is lacking. 
The decrease of the near-side-peak correlation is because that leading two jets are brought to insensitive regions around poles of the reference thrust axis. As a result of balance for the event thrust calculation, track pairs amongst on-axis jets tend to yield larger $\Delta\phi$ angular differences.
Compared to collisions at high center-of-mass energies, jets are composed of fewer constituents and have broader shapes at the low energy regime. This makes it hard to form a significant near-side-peak correlation.
We calculated the magnitude of the near-side-peak correlation with respect to different collision energies with \textsc{sherpa 2.2.5}~\cite{Sherpa:2019gpd} simulation of $e^+e^- \to $ hadrons at the leading order, and found results suggesting a significant correspondence.

%% plot -- 2PC (Data, 12-999 or 14-999) beam+thrust conti

%%%%%%%%%%%%%%%%%%%% ZYAM (W/ GENERATOR) %%%%%%%%%%%%%%%%%%%%
Evidence for the ridge signal can be best examined in the azimuthal differential yield $Y(\Delta \phi)$ by averaging the correlation function over the long-range region with $1.5 \le |\Delta \eta| < 3.0$.
A ``zero yield at minimum''(ZYAM) method~\cite{Ajitanand:2005jj} is further implemented to separate any enhanced near-side correlation around $\Delta \phi = 0$ distinct from a constant correlation.
The constant contribution along $\Delta \phi$, denoted as $C_{\rm ZYAM}$, is estimated by the minimum of the fit with a third-order Fourier series to the data points.
% ridge yield systematics
A fit with a third-order polynomial plus a cosine term, and with a fourth-order polynomial are also checked to estimate $C_{\rm ZYAM}$ in parallel. 

%%%%%%%%%%%%%%%%%%%% (SYST) %%%%%%%%%%%%%%%%%%%%
The systematic uncertainties due to the selection and correction operations 
% on the correlation function measurement 
are calculated with respect to  long-range $Y(\Delta \phi)$. 
% event selection
Hadronic-event selection is examined by tightening the energy sum requirement in the ECL from $E_{\rm sum}>0.18\sqrt{s}$ to $0.23\sqrt{s}$.
The primary-particle selection systematic is estimated by making variations of the proper lifetime requirement $\tau<1$ cm/$c$ and the vertex $V_{r}<1$ cm.
In general, results from both variations differ by less than 1\% (or by $1$-$6\%$ for the high-multiplicity bin).
% tracking
A 0.35\% uncertainty is quoted for the track reconstruction efficiency, which is evaluated by comparing partially and fully reconstructed $D^* \to \pi_{\text{slow}} D^0 (\to \pi^+ \pi^- K_S^0 (\to \pi^+ \pi^-))$ decays~\cite{BaBar:2014omp}.
% dominant systematics
In the beam axis analysis, the systematic uncertainties are mainly from the primary-particle selection and the tracking efficiency, which are in general on the order of 0.3\%--0.4\%, while the primary-particle selection increases to a 6.3\% uncertainty for high-multiplicity events with ${\rm N}_{\rm trk}^{\rm rec} \ge 14$. 
For high-multiplicity event classes in the thrust axis analysis, dominant sources of systematic uncertainties are due to the event selection ($<2\%$) and the primary-particle selection ($<4\%$), where the estimation of uncertainties suffers from the need for large statistics to derive a precise reweighting factor for the efficiency correction and for the mixed events. On the other hand, for low-multiplicity classes, the dominant source of uncertainty is due to tracking.
% else
Other uncertainties originate from MC reweighting, the $B(0,0)$ factor, mixed events reweighting, scaling corrections due to bin effects and residual bin effects, all of which are checked to be small in different multiplicity bins (of order $\mathcal{O}(10^{-4})$), with the largest one contributing up to 0.3\% uncertainty. 
% Detailed systematics uncertainties for different multiplicity intervals will be reported in Ref.~\cite{ToBePublished}.

\begin{figure}[ht]
\centering
		\includegraphics[width=0.48\textwidth]{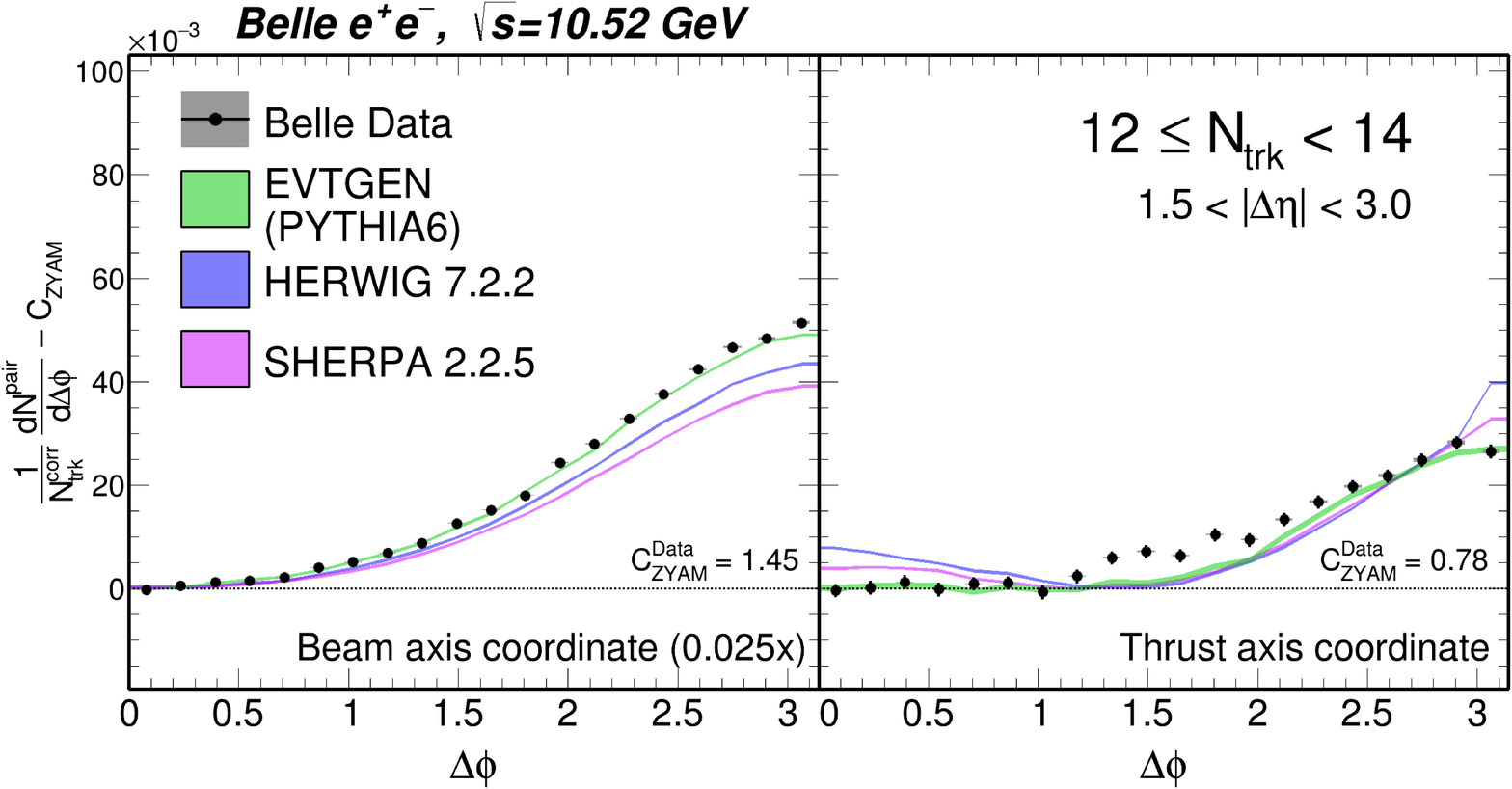}
		\includegraphics[width=0.48\textwidth]{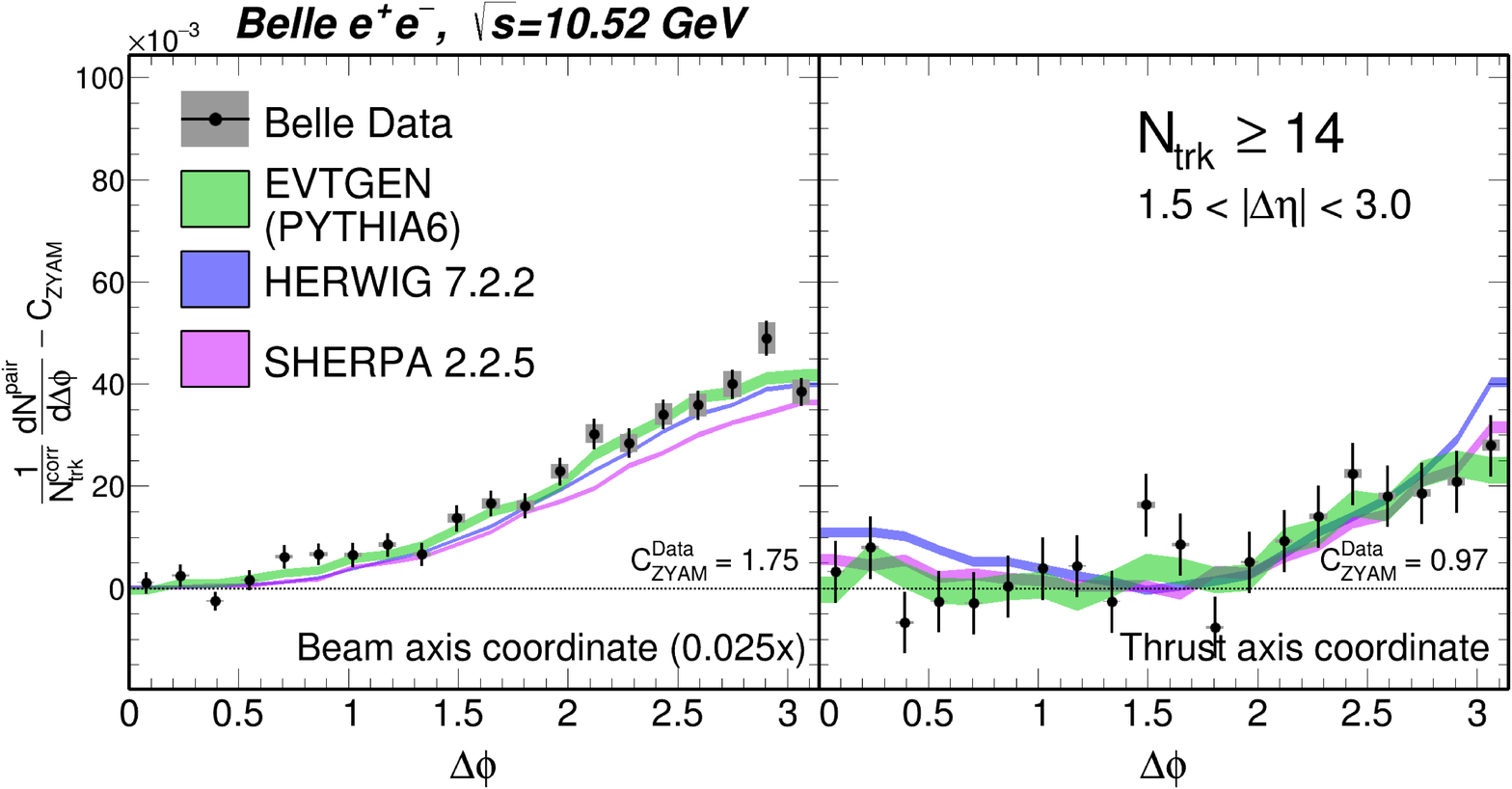}
\caption{Comparison of ZYAM-subtracted $Y(\Delta \phi)$ in the range $1.5 \le |\Delta \eta| < 3.0$ for the multiplicity $12 \le {\rm N}_{\rm trk}^{\rm rec} < 14$ (top) and ${\rm N}_{\rm trk}^{\rm rec} \ge 14$ (bottom), where subtracted constants of data $C_{\rm ZYAM}^{\rm Data}$ are quoted.
In each panel, results for beam (left) and thrust (right) axis analyses are displayed. The colored bands show simulation predictions from Belle MC (green), \textsc{herwig} (blue), and \textsc{sherpa} (violet). The error bars on the data represent the statistical uncertainties, and the gray boxes are systematic uncertainties. For visual purposes, the minimal statistical uncertainty of the MC colored bands is set to be $3 \times 10^{-4}$, and beam-axis ZYAM-subtracted yields are scaled by a factor of $0.025$.}
\label{fig:dNdphi_generator}
\end{figure}

Figure~\ref{fig:dNdphi_generator} shows the measurement of long-range $Y(\Delta \phi)$ after performing the ZYAM method, along with a comparison of predictions from Belle MC, \textsc{herwig 7.2.2}~\cite{Bellm:2015jjp}, and \textsc{sherpa 2.2.5}~\cite{Sherpa:2019gpd} event generators.
The region with small azimuthal angle difference ($\Delta \phi \approx 0 $) is where possible ridge signals would be visible as a nonzero value.
% beam axis coordinate
In the beam axis coordinates, all generators are consistent  with data in the near-side ridge region, but \textsc{herwig} and \textsc{sherpa} undershoot the data in the away-side region.
% thrust axis coordinate
In the thrust axis analysis,
the Belle simulation, with specific tunes adapted to Belle data, gives again a better description of these correlation data. A larger discrepancy from the data is seen in the \textsc{herwig} simulation. An excess of correlations is showing up in the near-side ridge-prone region and there is also an overshoot in the away-side region.

% Summary generators

%%%%%%%%%%%%%%%%%%%% RST (UL) AND SUMMARY %%%%%%%%%%%%%%%%%%%%

\begin{figure}[ht]
\centering
    \includegraphics[width=0.5\textwidth]{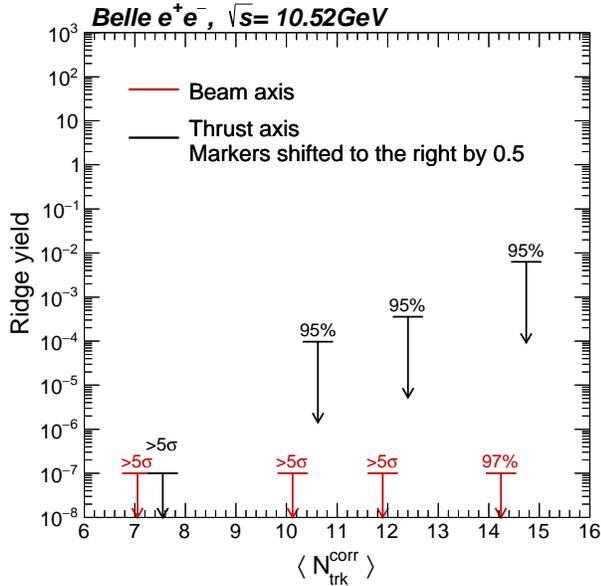}
\caption{Upper limits of the ridge yield as a function of $\langle {\rm N}_{\rm trk}^{\rm corr}\rangle$ in beam axis (red) and thrust axis (black) coordinate frames. Thrust axis limits are shifted to the right by 0.5 for presentation purposes. The label ``$>5\sigma$'' indicates the $5 \sigma$ confidence level upper limit.}
\label{fig:CLplots}
\end{figure}

The significance of any ridge signal can be quantified by integrating over $Y(\Delta\phi)$ from $\Delta\phi=0$ to where the ZYAM fit minimum occurs. 
Ridge yields smaller than an order of $10^{-10}$ are measured. Since there is no obvious ridge-like structure in either the beam or thrust axis analysis, a bootstrap procedure~\cite{Efron:1979bst} is implemented and the confidence limit of the integrated ridge yield is reported. In the bootstrap procedure, each azimuthal differential yield distribution is varied according to their statistical and systematic uncertainties, and the yield distribution is sampled $2\times10^6$ times.
For the ZYAM subtraction, three fit templates (a third-order Fourier series, a third-order polynomial plus a cosine term, and a fourth-order polynomial) are attempted, of which the most conservative confidence limit is quoted.
In Fig.~\ref{fig:CLplots}, the 95\% confidence level upper limits as a function of $\langle {\rm N}_{\rm trk}^{\rm corr}\rangle$ are reported. 
For the ridge yield upper limit less than $10^{-7}$, we report the confidence levels of the ridge signal exclusion, instead.

% CL plot

% measurement with Belle data
In summary, the first measurement of two-particle correlations of hadrons in beam and thrust axis coordinate systems, performed using $e^+e^-$ collision data at $\sqrt{s}= 10.52$ GeV from Belle is reported. 
A strong exclusion of ridge yield in the beam axis coordinate analysis is set. 
In the thrust axis coordinate analysis, there is no significant near-side-peak correlation, different from what was observed in hadronic collisions or high-energy $e^+e^-$ collisions. 
A 95-97\% confidence level upper limit or a $5 \sigma$ exclusion is set versus $\langle {\rm N}_{\rm trk}^{\rm corr}\rangle$  for the absence of any ridge yield in our measurement.
Though there is no hint of collectivity signals in the low-energy $e^+e^-$ collision system, the measurement can be provided as a reference for tunes of fragmentation models in the soft QCD scale.
% MC generator
Belle MC samples based on \textsc{evtgen}(with a \textsc{pythia6} interface for $e^+e^- \to q\bar{q}$ generation), along with \textsc{herwig} and \textsc{sherpa} event generators are examined.
Similar to the conclusion from the previous analysis with ALEPH archived data~\cite{Badea:2019vey}, the results in this study are found to be better described by \textsc{pythia6} than \textsc{sherpa} or \textsc{herwig}.

\vspace{0.3cm}

%***** Acknowledgments *****

%----------- Long version, for most papers ----------- 
We thank the KEKB group for the excellent operation of the
accelerator; the KEK cryogenics group for the efficient
operation of the solenoid; and the KEK computer group, and the Pacific Northwest National
Laboratory (PNNL) Environmental Molecular Sciences Laboratory (EMSL)
computing group for strong computing support; and the National
Institute of Informatics, and Science Information NETwork 5 (SINET5) for
valuable network support.  We acknowledge support from
the Ministry of Education, Culture, Sports, Science, and
Technology (MEXT) of Japan, the Japan Society for the 
Promotion of Science (JSPS), and the Tau-Lepton Physics 
Research Center of Nagoya University; 
the Australian Research Council including grants
DP180102629, % Sevior
DP170102389, % Varvell
DP170102204, % Yabsley
DP150103061, % Urquijo
FT130100303; % Urquijo;
Austrian Federal Ministry of Education, Science and Research (FWF) and
FWF Austrian Science Fund No.~P~31361-N36;
the National Natural Science Foundation of China under Contracts
No.~11435013,  %Zhen-An Liu
No.~11475187,  %Chang-Zheng Yuan
No.~11521505,  %Chang-Zheng Yuan
No.~11575017,  %Cheng-Ping Shen
No.~11675166,  %Wen-Biao Yan
No.~11705209;  %Yi-Ming Li
Key Research Program of Frontier Sciences, Chinese Academy of Sciences (CAS), Grant No.~QYZDJ-SSW-SLH011; % Chang-Zheng Yuan
the  CAS Center for Excellence in Particle Physics (CCEPP); %Chang-Zheng Yuan,
the Shanghai Science and Technology Committee (STCSM) under Grant No.~19ZR1403000; %Xiaolong Wang
the Ministry of Education, Youth and Sports of the Czech
Republic under Contract No.~LTT17020;
Horizon 2020 ERC Advanced Grant No.~884719 and ERC Starting Grant No.~947006 ``InterLeptons'' (European Union);
the Carl Zeiss Foundation, the Deutsche Forschungsgemeinschaft, the
Excellence Cluster Universe, and the VolkswagenStiftung;
the Department of Atomic Energy (Project Identification No. RTI 4002) and the Department of Science and Technology of India; 
the Istituto Nazionale di Fisica Nucleare of Italy; 
National Research Foundation (NRF) of Korea Grant
Nos.~2016R1\-D1A1B\-01010135, 2016R1\-D1A1B\-02012900, 2018R1\-A2B\-3003643,
2018R1\-A6A1A\-06024970, 2019K1\-A3A7A\-09033840,
2019R1\-I1A3A\-01058933, 2021R1\-A6A1A\-03043957,
2021R1\-F1A\-1060423, 2021R1\-F1A\-1064008;
Radiation Science Research Institute, Foreign Large-size Research Facility Application Supporting project, the Global Science Experimental Data Hub Center of the Korea Institute of Science and Technology Information and KREONET/GLORIAD;
the Polish Ministry of Science and Higher Education and 
the National Science Center;
the Ministry of Science and Higher Education of the Russian Federation, Agreement 14.W03.31.0026, % from 15.02.2018
and the HSE University Basic Research Program, Moscow; % from 15.04.2021
University of Tabuk research grants
S-1440-0321, S-0256-1438, and S-0280-1439 (Saudi Arabia);
the Slovenian Research Agency Grant Nos. J1-9124 and P1-0135;
Ikerbasque, Basque Foundation for Science, Spain;
the Swiss National Science Foundation; 
the Ministry of Education and the Ministry of Science and Technology of Taiwan;
and the United States Department of Energy and the National Science Foundation.

\bibliographystyle{apsrev}
\bibliography{belle2}

\end{document}